\documentclass[12pt]{article}

\usepackage{times,mathptm,amsmath}
\usepackage{sectsty}
\usepackage{balance}
\usepackage{setspace}
\usepackage{graphicx} 
\usepackage{float}

\author{Khadga J. Karki$^{\dagger}$, Fei Ma$^{\dagger}$, Kaibo Zheng$^{\dagger}$, Karel Zidek$^{\dagger}$\\
Abdelrazek Mousa$^{\dagger}$, Mohamed A. Abdellah$^{\dagger}$, Maria Messing$^{\ddagger}$\\
 L. Reine Wallenberg$^{\star}$, Arkadi Yartsev$^{\dagger}$, Tonu Pullerits $^{\dagger}$ \\
$^{\dagger}$\emph{\small Department of Chemical Physics, Lund University, Box 124, 22100 Lund, Sweden.} \\
$^{\ddagger}$ \emph{\small Solid State Physics, Lund Univerisity, Box 118, 22100 Lund, Sweden}\\
$^{\star}$\emph{\small Center for Analysis and Synthesis/nCHREM, Lund University, Box 124, 22100, Lund, Sweden}\\
\emph{\small Tel: +46 46 22 28131, Email: Khadga.Karki@chemphys.lu.se}}

\title{Multiple exciton generation in nano-crystals revisited: 
Consistent calculation of the yield based on pump-probe spectroscopy}

\begin{document}

\maketitle

\begin{abstract}
Multiple exciton generation (MEG) is a process in which more than one exciton is generated upon the absorption of a high energy photon, typically higher than two times the band gap, in semiconductor nanocrystals. It  can be observed experimentally using time resolved spectroscopy such as the transient absorption measurements. Quantification of the MEG yield is usually done by assuming that the bi-exciton signal is twice the signal from a single exciton. Herein we show that this assumption is not always justified and may lead to significant errors in the estimated MEG yields. We develop  a methodology to determine proper scaling factors to the signals from the transient absorption experiments. Using the methodology we find modest MEG yields in lead chalcogenide nanocrystals including the nanorods.

\end{abstract}

\doublespacing

MEG via impact ionization, enhanced by discrete energy levels due to quantum confinement in semiconductor quantum dots (QDs), was propsed by Nozik in 2001.~\cite{NOZIK2001} The process was experimentally observed by Klimov et al. in 2004 using pump-probe spectroscopy.~\cite{KLIMOV2004} 
Since then many researchers have investigated MEG in different QD materials using similar techniques, however, the yields have been found to be modest and the initial estimations of very high yields were attributed to the signal distortion due to the photocharging of the quantum dots and the surface defects.~\cite{BAWENDI2011} Though these controversies have been rectified in recent experiments, the reported MEG yields still diverge significantly.~\cite{NOZIK2010,GAFFNEY2009}  Bawendi and coworkers have raised the issue that the photo-luminescence intensity of QDs is not proportional to the number of excitons in the system and  proper scaling factors have to be used to quantify MEG yields in such experiments.~\cite{BAWENDI2007, BAWENDI2008} As far as we know, all the MEG studies using pump-probe spectroscopy have so far assumed that the observed signal scales linearly with the number of excitons. The assumption can be related to the simple state-filling argumentation where the bleach signal is due to to the gradual filling of the low lying conduction band electron states that correspond to the band-edge transitions. This qualitative assumption does not consider any refinement due to the confinement related correlation effects in QDs and might be insufficient for quantitative analyses. In this article we describe a refined analysis of the transient absorption (pump-probe)  measurements analogously to the photo-luminescence measurements.  We derive formulas to determine the calibration constants for the transient absorption measurements, and use them to calculate MEG yield in lead chalcogenide nano-crystals. Our calculations show that the photobleach signal due to a biexciton in nano-crystals is not always equal to twice the signal from a single exciton. Hence, proper scaling factors need to be determined for accurate calculation of the MEG yield. Though the methodology is general and applies to all the techniques that use time resolved spectroscopy with sub-picosecond time resolution, we focus on the time resolved bleach signal measured in our experiments.  

The physical processes that lead to the relaxation of bi- or multi-excitons are different to that of the single excitons;  Auger recombination dominates in the former while spontaneous emission plays the major role in the later. The time scales of Auger recombination --few tens of picoseconds in case of samples investigated here -- is distinctly different from that of the spontaneous emission --tens of nanoseconds. Consequently, the signatures of the single and multi-excitons can be conveniently distinguished in a time-resolved spectroscopy measurement as demonstrated below.~\cite{KLIMOV2004,BAWENDI2000}

\begin{figure}[htbp]
\includegraphics[width=3in]{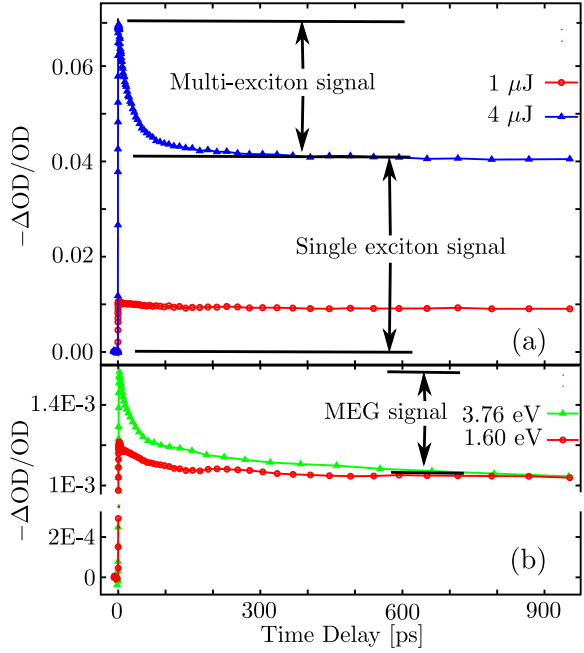}
\caption{ (a) Transient absorption signal from PbS QDs with band gap energy $E_g\approx 1.07$ eV  probed at $E_{probe}\approx 1.18$ eV when excited with laser pulses with photon energies $E_{ex}\approx 1.6$ eV. The signal at high pump intensity (blue curve, 4 $\mu$J energy per pulse) shows fast initial decay, which is due to the Auger recombination of the excitons when more than one exciton is created per QD. As the pump intensity is lowered (red curve, 1 $\mu$J energy per pulse), the probability that a QD absorbs more than one photon becomes negligible, consequently no more than one exciton is generated per QD and the fast decay vanishes. (b) Transient absorption signal when excited with high energy photons, $E_{ex} \approx 3.76 $ eV, (green curve) compared with the signal with low energy photon excitation. }
\label{MSIG}
\end{figure}

Figure \ref{MSIG}(a) shows the bleach signal of the probe pulse ( probe photon energy, $E_{probe} \approx 1.18 $ eV) in PbS quantum dot ($E_g\approx 1.07$ eV) under different pump intensities ( excitation photon energy, $E_{ex} \approx 1.6$ eV). At low pump intensity, when the probability of sequential absorption of two photons and hence the population of multi-excitons is negligible, the bleach signal (red line) does not show appreciable decay after the action of the pump. While at high pump intensities, when more than one exciton are created in some QDs, a fast decay within few tens of picoseconds is observed (blue line). The fast decay in the signal is due to the loss of the exciton population by Auger recombination. When the photon energy of the pump pulse is increased to values higher than $ 3 E_g$  the fast decay of the signal persists even for pump fluence for which the absorption of two photons can be safely neglected. Figure \ref{MSIG}(b) compares the bleach signal of the probe pulse when the samples are excited with pump-pulses with two different photon energies: the first with photon energy 1.6 eV where $E_{ex} < 2 E_g$ (red curve) and  the second with photon energy 3.76 eV where $E_{ex} > 3 E_g$ (green curve). The red curve in the figure has been multiplied by 0.82 to match the signal at long time delay with the green curve.  The average number of photons absorbed per quantum dots ( calculated using Equ.\eqref{EQ6}) in both excitations is less than 0.065. The prominent fast decay in the green curve is explained by the loss of multi-exciton population by Auger recombination as seen in the blue curve in Figure \ref{MSIG}(a); here multi-exciton is generated by MEG. Note that we use the term MEG explicitly to the process whereby the multi-excitons are generated by the splitting of a high energy single exciton to two or more low energy excitons rather than the sequential generation of excitons by the absorption of more than one photon. Though the amplitudes of the decay signals are the signatures of multi-excitons they are only a spectroscopic signatures of MEG.

The amplitudes of the signal due to different exciton numbers -- bi, tri, etc. -- in Figure \ref{MSIG} can be found by fitting the data with multiple exponentials;~\cite{BAWENDI2000} higher multi-excitons decay faster than the lower ones.  Quantifying the multi-exciton populations from the amplitudes, however, is non-trivial as the impact of complex electronic structure~\cite{ZUNGER2006,ZUNGER2007} on the dynamics and the amplitudes cannot be predicted \emph{per se}. Thus a proper scaling factor is necessary to compute  the multi-exciton population. In the following we describe how the scaling factors can be calculated from the time-resolved experiments themselves.

The probe signal immediately after the pump excitation with photon energies less than the MEG threshold can be written as:
\begin{equation}\label{EQ1}
S(t_0)=\sum_{i=1} a_i P(i)
\end{equation}
where $S(t)$ denotes the time varying signal, $t$ denotes the delay time with the subscript indicating the delay time between the pump and the probe pulse, $a_i$ is the signal contribution due to the $i^{th}$ exciton and $P(i)$ denotes the probability of $i^{th}$ exciton being populated by the absorption of $i$ photons. As usual, we use Poisson distribution:
\begin{equation}\label{EQ2}
P(i;\rho) = \frac{\exp(-\rho)\cdot \rho^i}{i!},
\end{equation}
where $\rho$  is the average number of photons absorbed per quantum dot. The probe signal is given by $S(t_0)=\rho \exp(-\rho)(a_1+a_2 \rho/2+\cdot\cdot\cdot$ .

For small values of $\rho$ ($\rho < 0.25$), the higher order terms can be neglected, and the signal can be written as:
\begin{equation}\label{EQ3}
S(t_0)=a_1 \rho \exp(-\rho)\left(\frac{k\rho+2}{2}\right),
\end{equation}
where $k=a_2/a_1$. 

The probe signal at long time delay, when all the bi-excitons have decayed to the single excitons, is $S(t_l)= a_1 \exp(-\rho)\sum_{i=1} \rho^i/i!$, which, when truncated after the second term, can be written as:
\begin{equation}\label{EQ4}
S(t_l)=a_1\rho\exp(-\rho)\left(\frac{\rho+2}{2}\right).
\end{equation}
 The relative strength of the biexciton contribution to the signal with respect to the single exciton contribution can be obtained from the ratio between the probe signal at the zero time delay and the long time delay:
\begin{equation}\label{EQ5}
k = \frac{x(\rho+2)-2}{\rho},
\end{equation} 
where $x=S(t_0)/S(t_l)$. $k$ is sufficient to quantify the MEG yield in the experiments where tri- or higher exciton formation by MEG can be neglected. Otherwise the higher order terms have to be included in Equ.\eqref{EQ3} and Equ.\eqref{EQ4} to determine even the calibration factors for the higher excitons (see the supporting information for the derivations of the equations in detail). 

The average number of photons absorbed per quantum dot, $\rho$, is related to the fluence: $\rho=\sigma_{abs} I$, where $\sigma_{abs}$ is the absorption cross-section at the pump wavelength. Using this relation in Equ.\eqref{EQ5} has drawbacks because $\sigma_{abs}$ is not always known, specially when new systems are investigated,~\cite{BONN2008} and the fluence $I$ can have large uncertainties unless one uses precisely calibrated detectors for specific wavelengths and accurately measures the focus spot size. To reduce the uncertainties we determine $\rho$ by fitting the fluence dependent long time delay transient absorption signal $S(I;t_l)$ to the following relation~\cite{TOMMY2012} (see the supporting information for the derivation):
\begin{equation}
S(I;t_l) = \alpha \left\{ 1- \exp\left(-\frac{I}{I_0}\rho_0\right)\right\},
\label{EQ6}
\end{equation}
where $\alpha$ is a constant, $I_0$ is the reference fluence and $\rho_0$ is the fitting parameter. This equation is particularly useful in experiments where the focus spot size of the pump beam is much bigger than the probe, like we are using here. When the probe focus is comparable to the pump, the bleach signal samples nano-particles that are excited with different intensities and the signals due to single and multi-excitons get averaged.~\cite{KARKI2012A} In this case a more generalized approach may be useful.~\cite{BAWENDI2008}

Note that Equ.\eqref{EQ6} uses the ratio of the fluences therefore any measurable quantity which is linearly related to the fluence can be used. In our experiments $I$ represents the voltage response from a photo-detector. Using linear response from the photo-detector substantially reduces the uncertainties related to the determination of the fluence.  In the following we use Equ.\eqref{EQ5} and Equ.\eqref{EQ6} to calculate the MEG yield in PbSe nanorods.

\begin{figure}[htbp]
\includegraphics[width=3in]{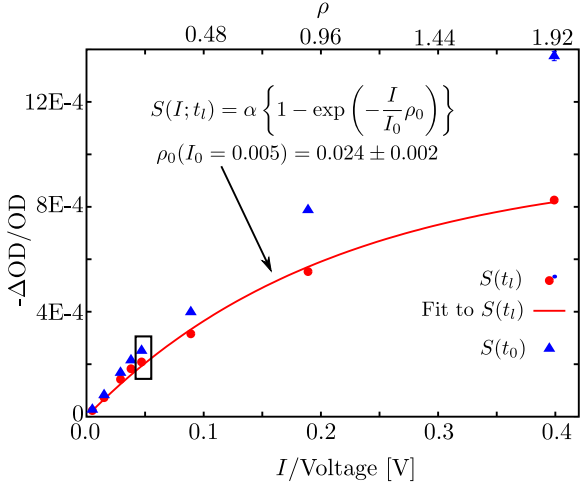}
\caption{Transient absorption signal of PbSe nanorods at different excitation intensities: blue points shortly after the pump excitation and red points after a long time delay. Red line is the fit to the long time signal using  Equ.\eqref{EQ6}. Upper axis is in the units of average excitation per QD.}
\label{STLTFIT}
\end{figure}

The nanorods were prepared using the protocol described by Melinger et al (the details of the sample preparation and the experimental setup can be found in the supporting information).~\cite{MELINGER2011} Figure \ref{STLTFIT} shows the probe signal ($\lambda=1100$ nm, $E_{probe}\approx 1.13$ eV) at short time delay, $t<15$ ps (blue points), and at long time delay, $t > 800 $ ps (red points), after the excitation with the pump pulse ($\lambda=775$ nm, $E_{ex}\approx 1.6 E_g$) for the different pump intensities. The $x$-axis in the figure is the voltage response from a photo-diode monitoring the pump pulses, which is linearly proportional to the intensity of the pump. The red line is the fit to the long time delay signal using Equ.\eqref{EQ6}, which gives $\rho_0(I_0=0.005 V)=0.024\pm0.002$; $\rho$ for other intensities can be calculated using the relation $\rho\propto I$.
We use the data point (boxed in Figure \ref{STLTFIT}) with $\rho=0.226$, $S(t_0)=2.51\times 10^{-4}\pm 1\times 10^{-6}$ and $S(t_l)= 2.08\times 10^{-4}\pm 3\times 10^{-6}$ to determine $k$, the relative contribution to the probe signal from biexcitons. For $\rho=0.226$, $P(2)=0.02$ and $P(3)=0.0015$, so that tri- and higher exciton contribution can be safely neglected while at the same time there is enough contribution from the biexcitons. Using Equ.\eqref{EQ5} we get $k=3.1\pm 0.4$; error propagation from the uncertainties in $x$, $\Delta x = 0.02$, and $\rho$, $\Delta \rho=0.002$, is used to compute the uncertainty in $k$. For other values of $\rho$, $0.25 > \rho > 0.1 $, we get $3.5 > k > 3.1$, which is within the uncertainty range of $k$ at $\rho=0.226$. For $\rho < 0.1 $ biexciton contribution is negligible while for $\rho > 0.25$ tri-exciton contribution cannot be neglected. Similar calculations done on PbS QDs using the data shown in Figure \ref{QDSTLT} give $k = 1.9\pm 0.2$. So the scaling factors are not the same for different nano-particles. They could as well depend on the wavelength of the light used in the experiments.

\begin{figure}[htpb]
\includegraphics[width=3.25in]{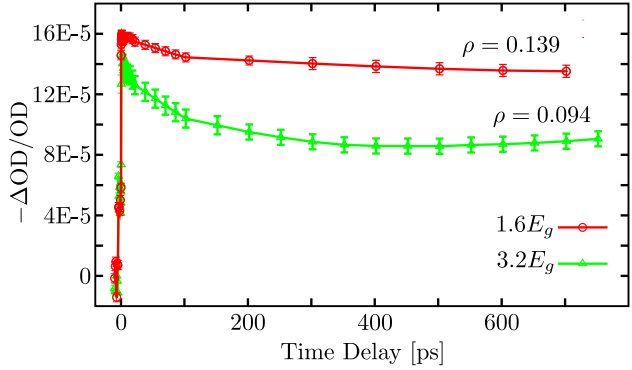}
\caption{ Time resolved probe signal for $1.6$ eV (red curve) and $3.2$ eV (green curve) pump-photon excitations. The band gap of the PbSe nanorods is about 1 eV. The green curve shows prominent fast initial decay even for lower average number of photons absorbed per quantum dot than for the red curve, which indicates MEG. The quantum yield of the MEG in this case is about 1.21.}
\label{NRMEG}
\end{figure}

\begin{figure}[h]
\includegraphics[width=3in]{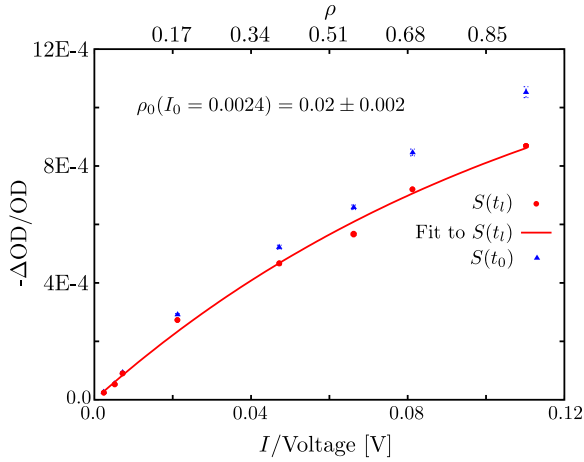}
  \caption{Bleach signal vs. intensity (photo-voltage from a linear detector) at two different delay times in PbS QDs. The red points show the probe signal  at long time, 800 ps after the pump excitation
  with various fluences. Red line is the fit to the data points using
  Equ.\ref{EQ6}. The blue points show the corresponding probe signal at short time, about 3 ps after the temporal overlap of the pump and the probe beams. Upper axis is in the units of average excitation per QD. }
\label{QDSTLT}
\end{figure}

Figure \ref{NRMEG} compares the probe signal when the PbSe nanorods are excited with $3.2 E_g$ (green curve; UV excitation) and $1.6 E_g$ (red curve; NIR excitation) photons. $\rho=0.139$ for the NIR excitation is calculated from the Figure \ref{STLTFIT} and the $\rho=0.094$ for UV excitation is calculated using the relation $S(t_l)\propto \rho$ for $\rho << 1$ (see Equ.\eqref{EQ4} and the supporting information for the details). The signals immediately and after long time delay following the UV excitation are $S(t_0)=(1.48\pm0.06)\times 10^{-4}$ and $S(t_l)=(0.95\pm0.04)\times 10^{-4}$, respectively. For $k=3.1$ and $\rho=0.094$ the expected signal  $S_{ex}(t_0)$  calculated using Equ.\eqref{EQ5} if there was no MEG would be $(1.04\pm0.06)\times 10^{-4}$. The rest of the signal, $S(t_0)-S_{ex}(t_0)=(0.44\pm0.08)\times 10^{-4}$, is due to the MEG.
If $\eta$ is the MEG yield, the fraction of the initially populated excitons that undergo MEG, then the total initial signal can be written as the sum of the signal from the bi-excitons, $\eta k S_{ex}(t_0 )$, and the single excitons, $(1-\eta)S_{ex}(t_0)$:
\begin{equation}\label{EQ7}
S(t_0) = \eta (k-1)S_{ex}(t_0)+S_{ex}(t_0).
\end{equation}
Using the values for $S(t_0)$, $S_{ex}(t_0)$ and $k$ in Equ.\eqref{EQ7}, we get $\eta=0.21\pm0.05$. The corresponding quantum yield of the exciton generation is $ \phi=1.21\pm 0.05$.  
 
The MEG yield, $\eta$, we have obtained in the NRs is slightly less than the yield in QDs but within the error range.~\cite{NOZIK2010} Our results agree with some previous works, which show that Auger like processes in NRs can be suppressed as compared to QDs.~\cite{KLIMOV2003} On the other hand, our results do not show significant enhancement in the MEG yield in NRs as observed recently.~\cite{MELINGER2011}  However, these conflicting results are not directly comparable as the previous calculations~\cite{MELINGER2011} implicitly assume $k=2$, which leads to different yields even if the data are similar.

As far as we know all of the work on the quantification of the MEG yields based on the pump-probe measurements on QDs use $k=2$. This value is based on the assumption that the exciton bleach signal is due to state filling of the exciton bands. In the case of PbS QDs, the lowest band can accomodate 8 excitons. According to the assumption, the signal $S(t_0)$ should scale linearly with intensity untill the lowest exciton band is completely filled with the excitons. However, even for modest intensities for which $\rho < 1$ the signal $S(t_0)$ shown in Figure \ref{STLTFIT} for the NRs and Figure \ref{QDSTLT} for the PbS QDs deviates from  the linear relationship. This indicates the possibility of exciton-exciton correlation effects in nano-particles in strong confinement regime. Though these effects have not been the focus of current research, further studies of such effects might be important for the deeper understanding of MEG in nano-particles. 

Recent studies argue that the observed MEG yields in QDs may not improve the efficiency of photo-voltaic cells as compared to conventional solid state devices.~\cite{BONN2009} Consequently, the current research trend has diverged into investigating nano-particles whose shape, size and composition are different from the idealized quantum dots. It has been observed that these modifications can dramatically alter the spectroscopic properties of the nano-particles.~\cite{KLIMOV2003} In this context assuming $k=2$ when quantifying the MEG yields cannot be justified. Moreover, the MEG yields determined by photoluminescence measurements~\cite{BAWENDI2008} that take into account the proper scaling factors differ from the yields obtained by the pump-probe techniques. Use of the proper scaling factors in the pump-probe measurements could provide valuable insight into the discrepancies.

To conclude, our experiments show only modest MEG yields in PbSe NRs. Using our methodology we found that the bleach signal due to a biexciton in nano-particles is not always twice the bleach signal from a single exciton. It is important to explicitly compute the proper scaling factors that relate the populations to the signal amplitudes for the accurate quantification of the MEG yields in different nano-particles. It is also important to follow such a methodology for meaningful comparison the MEG yields obtained by using different experimental techniques.

\providecommand*\mcitethebibliography{\thebibliography}
\csname @ifundefined\endcsname{endmcitethebibliography}
  {\let\endmcitethebibliography\endthebibliography}{}

\textbf{Acknowledgement}

We gratefully acknowledge financial support from the Swedish Research Council, the Knut and Alice Wallenberg Foundation, the Wenner-Gren Foundations, Crafoord foundation, Swedish Energy Agency and the Swedish Foundation for Strategic Research. We thank Torbj\"orn Pascher for the help with the time-resolved experiments. Collaboration within nmC@LU is acknowledged.

\textbf{ Supplementary Information}

\textbf{Experimental setup}

Figure below shows the schematic of the experimental setup.

\begin{figure}[H]
  \includegraphics[width=5in]{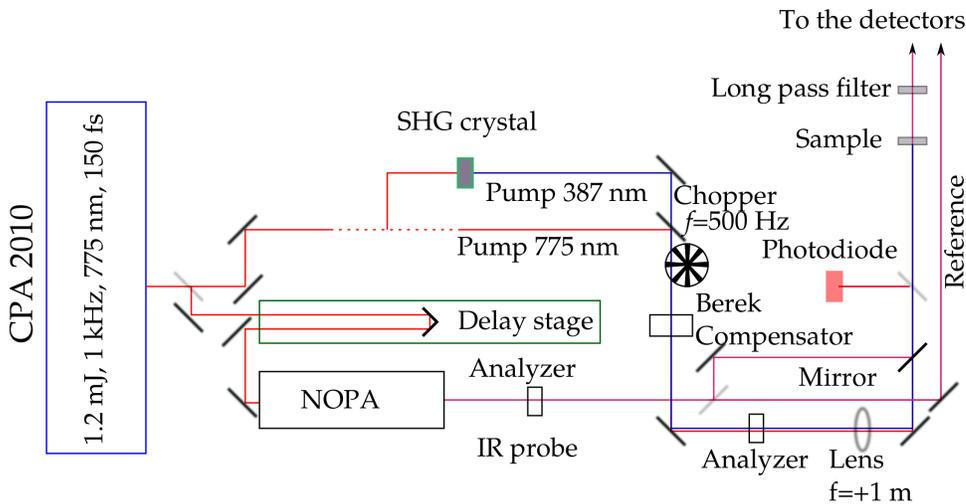}
  \caption{Schematic of the pump-probe setup used in the transient absorption
  measurements.}
\end{figure}

The pulses (1.2 mJ pulse energy, 1 kHz repetition rate, center wavelength at
775 nm and pulse duration about 150 fs) generated from the amplified laser \
(CPA 2010) are split using a beam splitter to generate the pump as well as the
probe pulse. The probe pulse in the infra-red region ($\lambda > 1000$ nm) are
generated using a non-collinear optical parametric amplifier (NOPA). An
analyzer is used to define the horizontal polarization of the pulses. The
delay stage before the NOPA controls the time delay between the pump and the
probe pulses. A 50 cm focal length concave mirror with a hole at the center is
used to focus the probe pulses to the sample. The 775 nm pulses from the
amplifier are used as the pump for the NIR pump (for the experiments done to
measure the Auger decay without MEG). The 387 nm pulses from the second
harmonic of the 775 nm pulses are used as the pump for the UV excitation to
investigate the MEG process. The pump pulses are choped at 500 Hz using a
mechanical chopper. A Berek compensator is used to rotate the polarization of
the pump beam and an analyzer is used to define the horizontal polarization.
The intensity of the pump beam is controlled by the compensator/analyzer
combination. The probe beam is polarized at the `magic angle' with respect to
the pump. The pump beam is focused into the sample using a lens with focal
length 1 m. A little fraction of the pump beam is directed to the
photo-dectector to monitor the intensity of the pump pulses. The pump diameter
at the focus is about 0.7 mm and the probe diameter is about 0.4 mm. The
optical density of the samples at the excitation wavelengths is kept below
0.3. At the absorption cross-sections at the different excitation wavelenghts
are different, the signals need to be normalized by the concentration of the
nano-crystals in the solution for the comparison. We have chosen to normalize
them by the optical density at 450 nm. The samples are shaked during the
measurements to avoid degradation due to photocharging. Typical error in the measurements is in the order of 10$^{-7} \Delta$ OD. At such low error condition the signatures of photocharges can be seen as the pre-pump signal. We see such pre-pump signal in the cases with high pump fluence where the signal is in the order of 10$^{-2} \Delta$ OD (shown in Fig. 2 ). All the measurements used in the analysis are done with pump fluence where the signal is at least an order of magnitude smaller (10$^{-3} \Delta $ OD). The signal level in our measurements for MEG is in the order of $10^{-5} \Delta$ OD, about three order of magnitude smaller than in the case when photocharging is discernible. No photocharging has been reported for the corresponding average number of excitons excited in the QDs. We probe above the band edge to avoid the artifacts due to ultrafast (sub-picosecond) de-population of the band edge by surface trapping. We also set $S(t_0)$ to 3 ps after the pump pulse to avoid the sub-picosecond de-population component, if present. 

\begin{figure}[H]
\includegraphics[width=12cm]{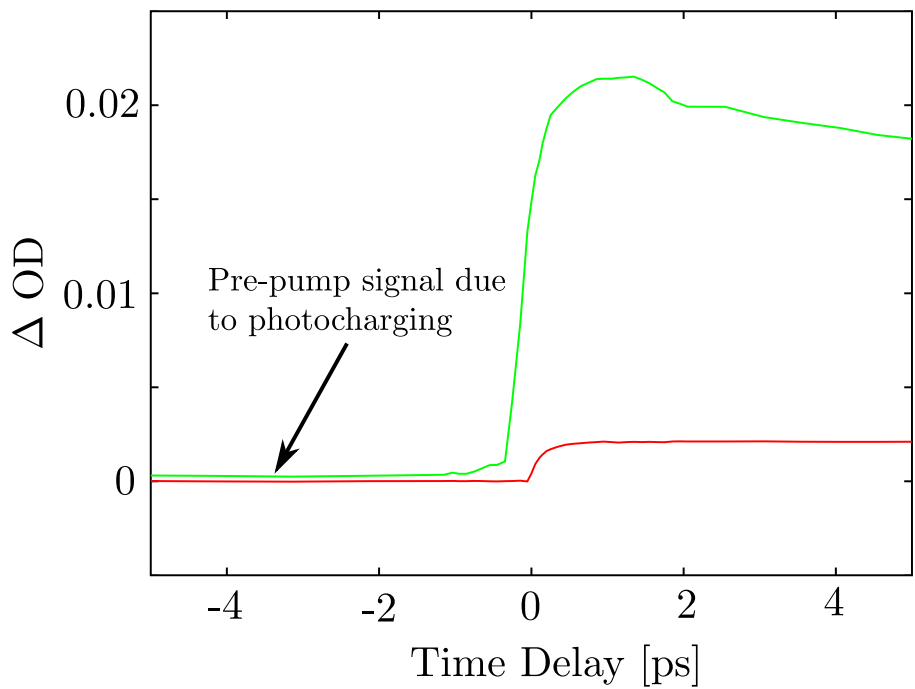}
\caption{Pre-pump signal due to photocharging of the nano-crystals under high pump fluence (gree curve). At low pump fluence such effect is not observed (red curve).}
\end{figure}

\textbf{Sample preparation}

{\textbf{Materials}}

Lead oxide ($> 99.9\%$), oleic acid (OA,$ >$99\%), selenium (99.5\%, 100 mesh),
trioctylphosphine (TOP,$>$90\%), tris(diethylamino)phosphine (TPD, 97\%) and
bis(trimethylsilyl) sulfide (TMS) were purchased from Aldrich. Solvents used
in the synthesis included 1-octadecene (ODE, 90\%), toluene (anhydrous,
99.8\%), hexane (95\%), heptane (96.7\%), acetone (HPLC, $>$99.8\%), chloroform
($>$99.9\%) and ethanol (99.7\%). All chemicals were used as received.

{\textbf{Synthesis}}

The synthesis method for the quantum dots can be found elsewhere.$^1$ Briefly,
446 mg (2 mmol) lead PbO was dissolved in a mixture of 2 ml OA and 20 ml ODE
then was heated to 90 $^0$C under N$_2$ in a three necked flask. After PbO is
completely decomposed to form colorless Pb oleate the solution further heated
to various temperatures for QDs growth. 0.2 ml of TMS dissolved in 10 ml ODE
was rapidly injected into the Pb oleate solution. The rapid injection of TMS
solution into the reaction flask changed the color of the reaction mixture
from colorless to deep brown. The reaction continued for 2 min. The solution
in the flask was dissolved in 10 ml of toluene and was precipitated with
methanol and acetone and then redispersed in non-polar solvent such as hexane
and toluene for storage. To vary the size of the nano-particles, the injection
temperature was changed. The TEM image of the QDs grown at 160 $^0$C are is
shown in Fig. 2.

\begin{figure}[htb]
  \includegraphics[width=3in]{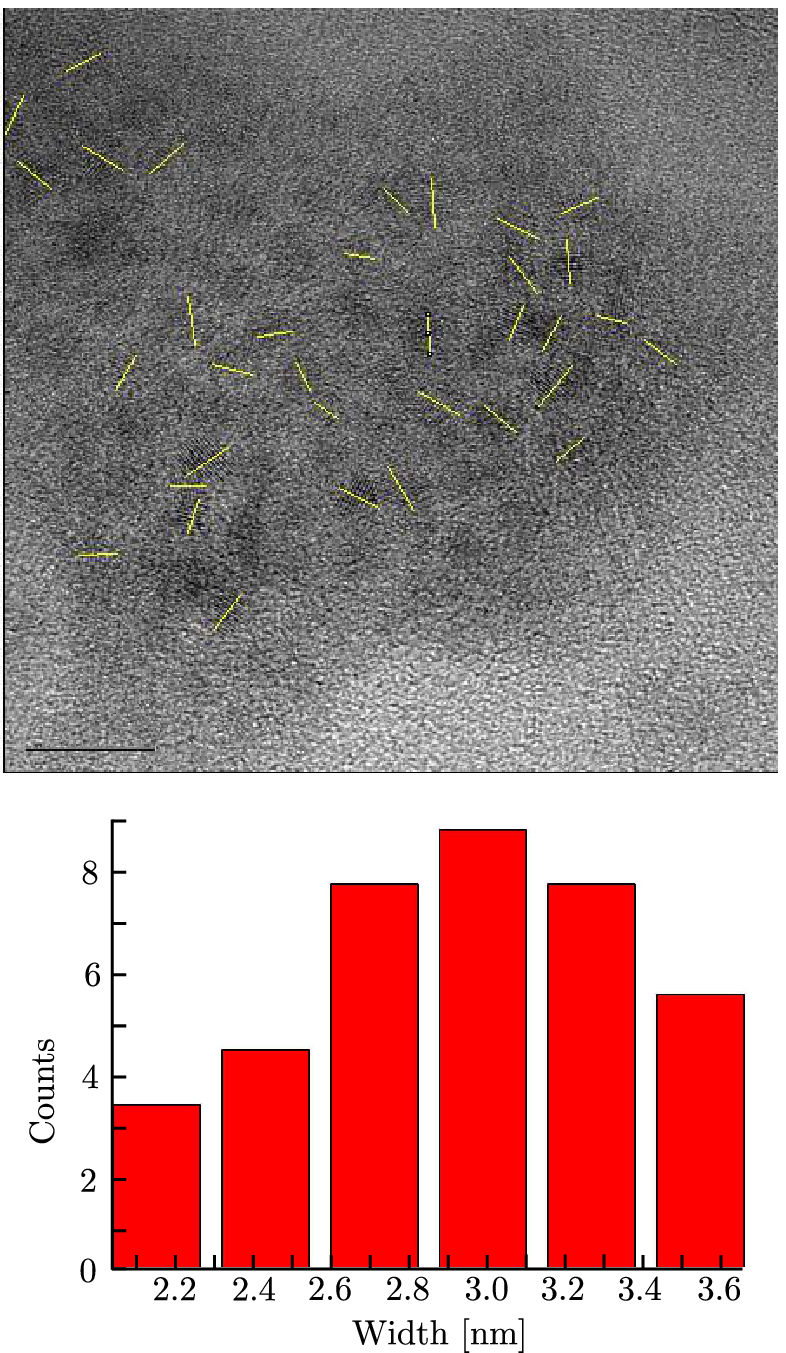}
  \caption{TEM image of the QDs. The average diameter is 3.1 nm.}
\end{figure}

For the synthesis of the nanorods 4 mmol PbO and 10 mmol OA were mixed with ODE with a total weight of
16 g in a three-neck flask. The mixture was then purged with N$_2$ and
heated to 150 $^0$ C to form colorless Pb oleate solution. Then 5.9 ml
Se-TDP solution with 6 mmol Se was injected into Pb oleate solution containing
2 mmol PbO, 6 mmol OA and 31 mmol ODE at 130 $^0$ C and allowed to
react for 2 mins. The aliquot was rapidly cooled using ice bath. Heptane and
ethanol were used to purify the NRs for at least twice and finally dissolved
in hexane. The morphology of the obtained nanorods is shown in Fig. 3.

\begin{figure}[htb]
\includegraphics[width=3in]{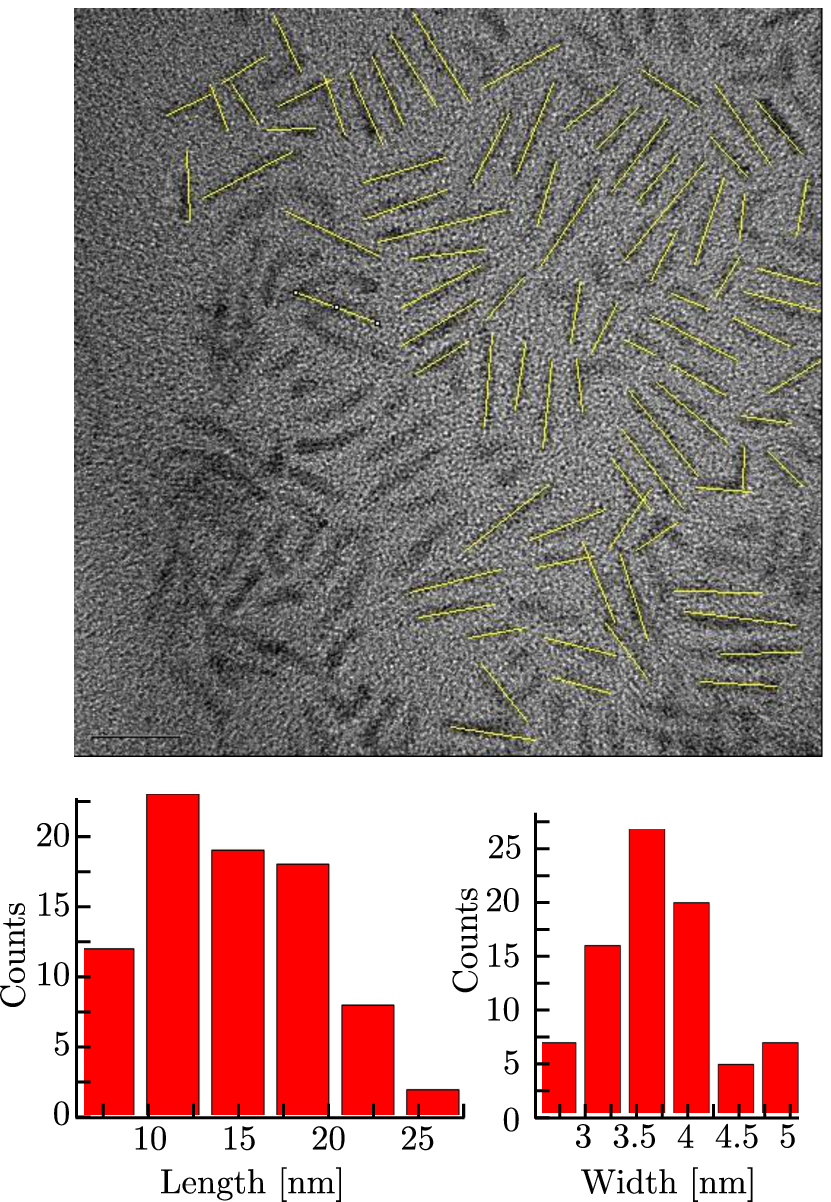}
  \caption{TEM image of the nanorods. The average length is 16.5 nm and the
  average width is 3.9 nm.}
\end{figure}

\textbf{Derivation of the equations}

{\textbf{Determination of $k \, , l \textrm{ and } \, m$ values}}.

The signal at the initial times is given by:
\begin{equation}
  S ( t_0) = \sum_{i = 1} a_i P ( i ; \rho) \label{EQ1}
\end{equation}
where $P ( i ; \rho)$ gives the probability of $i$ photons being absorbed by a
nano-particle given that the average number of photons absorbed is $\rho$;
here the Poisson distribution is used:
\begin{equation}
  P ( i ; \rho) = \frac{\exp ( - \rho) \cdot \rho^i}{i!} . \label{EQ2}
\end{equation}
Then Eq.(\ref{EQ1}) can be written as
\begin{eqnarray}
  S ( t_0) & = & \exp ( - \rho) \left\{ \frac{a_1 \rho}{1} + \frac{a_2
  \rho^2}{2!} + \frac{a_3 \rho^3}{3!} + \frac{a_4 \rho^4}{4!} + \ldots
  \right\} .  \label{EQ3}
\end{eqnarray}
When absorption of more than two photons can be neglected, Eq.(\ref{EQ3}) can
be truncated to the second order term:
\begin{eqnarray}
  S ( t_0) & = & \exp ( - \rho) \left\{ \frac{a_1 \rho}{1} + \frac{a_2
  \rho^2}{2!} \right\} = \rho \exp ( - \rho) \left\{ a_1 + \frac{a_2 \rho}{2}
  \right\} = \rho \exp ( - \rho) \left\{ \frac{2 a_1 + a_2 \rho}{2} \right\}
  \nonumber\\
  & = & a_1 \rho \exp ( - \rho)  \left\{ \frac{2 + \rho a_2 / a_1}{2}
  \right\} = a_1 \rho \exp ( - \rho) \left\{ \frac{k \rho + 2}{2} \right\}
  \,\,, \textrm{with} k = a_2 / a_1 .  \label{EQ4}
\end{eqnarray}
The signal after long time delay, $t_l$, is given by:
\begin{eqnarray}
  S ( t_l) & = & a_1 \sum_{i = 1} P ( i ; \rho) = a_1 \exp ( - \rho)  \left\{
  \rho + \frac{\rho^2}{2!} + \frac{\rho^3}{3!} + \frac{\rho^4}{4!} + \ldots
  \right\},  \label{EQ5}
\end{eqnarray}
which when truncated after the second term in the sum becomes
\begin{eqnarray}
  S ( t_l) & = & a_1 \rho \exp ( - \rho)  \left\{ \frac{2 + \rho}{2} \right\} 
  \label{EQ6}
\end{eqnarray}
The ratio of the short time and long time signal gives (from Eq.(\ref{EQ4})
and Eq.(\ref{EQ6}))
\begin{eqnarray}
  \frac{S ( t_0)}{S ( t_l)} & = & x = \frac{k \rho + 2}{\rho + 2} \nonumber\\
  \Rightarrow k & = & \frac{x ( \rho + 2) - 2}{\rho}  \label{EQ7}
\end{eqnarray}
When the absorption of {\textbf{three photons}} cannot be neglected, we need
to use the following equations for calculating the signal contributions from
the two and three photon absorptions:
\begin{eqnarray}
  S ( t_0) & = & \left( \frac{a_3 \rho^3}{6} + \frac{a_2 \rho^2}{2} + a_1 \rho
  \right) \exp ( - \rho) = \left( \frac{l \rho^2}{6} + \frac{k \rho}{2} + 1
  \right) a_1 \rho \exp ( - \rho) \nonumber\\
  S ( t_l) & = & \left( \frac{\rho^2}{6} + \frac{\rho}{2} + 1 \right) a_1 \rho
  \exp ( - \rho)  \label{EQ8}
\end{eqnarray}
with $l = a_3 / a_1$ and $k = a_2 / a_1$,

and
\begin{eqnarray}
  \frac{S ( t_0)}{S ( t_l)} & = & x = \frac{l \rho^2 + 3 k \rho + 6}{\rho^2 +
  3 \rho + 6} \nonumber\\
  \Rightarrow l \rho^2 + 3 k \rho - ( x \rho^2 + 3 x \rho + 6 ( x - 1)) & = &
  0.  \label{EQ9}
\end{eqnarray}
Eq.(\ref{EQ9}) is a linear equation with two variables $l \textrm{ and }\, k$ and it
can be solved by using two data points with different $\rho \textrm{ and } x$
values ($\rho_1 $, $\rho_2$, $x_1$ and $x_2$, respectively).
The solutions are
\begin{eqnarray}
  l & = & \frac{\rho_2 R - \rho_1 S}{T} \nonumber\\
  k & = & - \frac{\rho_2^2 R - \rho_1^2 S}{3 T}  \label{EQ10}
\end{eqnarray}
with $R = x_1  ( \rho_1^2 + 3 \rho_1 + 6) - 6$, $S = x_2  ( \rho_2^2 + 3
\rho_2 + 6) - 6$ and $T = \rho_1 \rho_2  ( \rho_1 - \rho_2)$.

Similarly if the absorption of {\textbf{four photons}} cannot be neglected,
the following formulas need to be used to calculate the signal contributions
from the two, three and four photons respectively.
\begin{eqnarray*}
  m & = & \frac{W T + V S + U R}{Z} \,\,, m = a_4 / a_1\\
  l & = & - \frac{( \rho_1 + \rho_2) W T + ( \rho_1 + \rho_3) V S + ( \rho_2 +
  \rho_3) U R}{4 Z}\\
  k & = & \frac{\rho_1 \rho_2 W T + \rho_1 \rho_3 V S + \rho_2 \rho_3 U R}{12
  Z},
\end{eqnarray*}
where $T = x_3  ( \rho_3^3 + 4 \rho_3^2 + 12 \rho_3 + 24) - 24$, $S = x_2  (
\rho_2^3 + 4 \rho_2^2 + 12 \rho_2 + 24) - 24$, $R = x_1  ( \rho_1^3 + 4
\rho_1^2 + 12 \rho_1 + 24) - 24$, $W = \rho_1 \rho_2 ( \rho_2 - \rho_1)$, $V =
\rho_1 \rho_3 ( \rho_1 - \rho_3)$, $U = \rho_2 \rho_3 ( \rho_3 - \rho_2)$ and
$Z = \rho_1 \rho_2 \rho_3 ( \rho_1 - \rho_2) ( \rho_2 - \rho_3) ( \rho_3 -
\rho_1)$.

{\textbf{Determination of $\rho$}}.

For the signal at long time delays, Eq.(\ref{EQ5}) can be simplified as
follows
\begin{eqnarray}
  S ( t_l) & = & a_1 \exp ( - \rho) ( \exp ( \rho) - 1) = a_1 ( 1 - \exp ( -
  \rho)) .  \label{EQ11}
\end{eqnarray}
The decay of the single exciton within the time delay, $t_l$, cannot be
neglected, the equation has to be modified to
\begin{eqnarray}
  S ( t_l) & = & \alpha ( 1 - \exp ( - \rho))  \label{EQ12}
\end{eqnarray}
with $\alpha = a_1 \exp ( - t_l / \tau_1)$, where $\tau_1$ is the life of the
single excitons. If the fluence, $I$, of the pulses is measured, then $\rho$
can be written as
\begin{eqnarray}
  \rho & = & \sigma_{abs} I  \label{EQ13}
\end{eqnarray}
and for a reference fluence, $I_0$
\begin{eqnarray}
  \rho_0 & = & \sigma_{abs} I_0 .  \label{EQ14}
\end{eqnarray}
From Eq.(\ref{EQ13}) and (\ref{EQ14}) we have
\begin{eqnarray}
  \rho & = & \frac{I}{I_0} \rho_0  \label{EQ15}
\end{eqnarray}
and substituting $\rho$ in Eq.(\ref{EQ12}) we get
\begin{eqnarray}
  S ( t_l) & = & \alpha \left\{ 1 - \exp \left( - \frac{I}{I_0} \rho_0 \right)
  \right\} .  \label{EQ16}
\end{eqnarray}
{\textbf{Note}} that for $\rho \ll 1$, Eq.(\ref{EQ12}) can be simplified to
(using first order approximation)
\begin{eqnarray}
  S ( t_l) & = & \alpha ( 1 - ( 1 - \rho)) = \alpha \rho \nonumber\\
  \Rightarrow S ( t_l) & \propto & \rho .  \label{EQ17}
\end{eqnarray}

\vspace{1.2in}

{\textbf{References}}

1. M. Hines et al., {\emph{Adv. Mater.}}, {\textbf{2003}}, 15, 1844-1849.

\end{document}